\documentclass[prd,preprint,tightenlines,floatfix,showpacs,preprintnumbers,nofootinbib,eqsecnum]{revtex4}
 \usepackage[dvips,final]{graphicx}
  \usepackage{amssymb}
   \usepackage{amsmath}
    \usepackage{amsfonts}
     \usepackage{epsfig}
      \usepackage{bm}

 \newcommand\beq{\begin{equation}}
 
 \newcommand\eeq{\end{equation}}
 \newcommand\beqn{\begin{eqnarray}}
 \newcommand\eeqn{\end{eqnarray}}

\begin{document}

\begin{flushright}
LU TP 13-07\\
February 2013
\end{flushright}

\title{Dark Energy from graviton-mediated
interactions\\ in the QCD vacuum}

\author{Roman Pasechnik}
\email{Roman.Pasechnik@thep.lu.se} \affiliation{Department of
Astronomy and Theoretical Physics, Lund University, SE-223 62 Lund,
Sweden}

\author{Vitaly Beylin}
\affiliation{Research Institute of Physics, Southern Federal
University, 344090 Rostov-on-Don, Russian Federation}

\author{Grigory Vereshkov}
\affiliation{Research Institute of Physics, Southern Federal
University, 344090 Rostov-on-Don, Russian Federation}
\affiliation{Institute for Nuclear Research of Russian Academy of
Sciences, 117312 Moscow, Russian Federation\vspace{1cm}}

\begin{abstract}
\vspace{0.5cm} Adopting the hypothesis about the exact cancellation
of vacuum condensates contributions to the ground state energy in
particle physics to the leading order in graviton-mediated
interactions, we argue that the observable cosmological constant can
be dynamically induced by an uncompensated quantum gravity
correction to them after the QCD phase transition epoch. To start
with, we demonstrate a possible cancellation of the quark-gluon
condensate contribution to the total vacuum energy density of the
Universe at temperatures $T<100$ MeV without taking into account the
graviton-mediated effects. In order to incorporate the latter, we
then calculate the leading-order quantum correction to the classical
Einstein equations due to metric fluctuations induced by the
non-perturbative vacuum fluctuations of the gluon and quark fields
in the quasiclassical approximation. It has been demonstrated that
such a correction to the vacuum energy density has a form
$\varepsilon_{\Lambda}\sim G \Lambda_{\rm QCD}^6$, where $G$ is the
gravitational constant, and $\Lambda_{\rm QCD}$ is the QCD scale
parameter. We analyze capabilities of this approach based on the
synthesis between quantum gravity in quasiclassical approximation
and theory of non-perturbative QCD vacuum for quantitative
explanation of the observed Dark Energy density.
\end{abstract}


\maketitle

\section{Introduction}

The existence of mysterious Dark Energy which drives current
accelerated expansion of the Universe is confirmed in many
cosmological observations so far, e.g. in studies of the type Ia
Supernovae \cite{SNE1A}, cosmic microwave background anisotropies
\cite{WMAP}, large scale structure \cite{SDSS} etc. The hypothesis
about the time-independent cosmological constant called
$\Lambda$-term is an essential part of the Standard Cosmological
Model known as the Cold Dark Matter with $\Lambda$-term Model (or
$\Lambda$CDM) and agrees well with all observational data collected
so far at the current level of experimental accuracy. The most
recent Planck 2013 data \cite{Ade:2013zuv} have not indicated any
dynamical (time-dependent) signatures of the Dark Energy thus
further supporting the $\Lambda$-term approximation. However,
theoretical origin of the cosmological constant has not been
properly understood yet.

Nowadays, the Dark Energy problem remains one of the major unsolved
problem of theoretical physics \cite{Weinberg:1988cp}. On the way of
searching for possible solutions of this problem many various
pathways were explored during last few decades referring to e.g. new
exotic forms of matter (e.g. ``quintessence'' \cite{quintessence},
``phantom'' \cite{phantom} etc), holographic models
\cite{holography}, string theory landscape
\cite{strings,Polchinski:2006gy}, Born-Infeld quantum condensate
\cite{Elizalde:2003ku}, modified gravity approaches
\cite{modgra,Bamba} etc. For a comprehensive overview of existing
theoretical models and interpretations of the Dark Energy, see e.g.
Refs.~\cite{Copeland:2006wr,Bamba:2012cp,Li:2012dt,Yoo:2012ug} and
references therein. In this work we are primarily focused on the
class of those approaches which are based upon conventional quantum
field theory and standard quantum gravity in quasiclassical
approximation.

The traditional identification of the $\Lambda$-term satisfying the
equation of state $P_{\Lambda}=-\varepsilon_{\Lambda}$ with vacuum
pressure $P_{\Lambda}$ and energy density $\varepsilon_{\Lambda}$
suffers from a large gap of knowledge on non-perturbative dynamics
of the ground state in particle physics. In particular, individual
vacuum condensates e.g. those which are responsible for the chiral
(quark-gluon condensate) and gauge (Higgs condensate) symmetries
breaking in the Standard Model contribute to the vacuum energy of
the Universe individually exceeding the observable value of the Dark
Energy by many orders of magnitude in absolute value
\cite{Martin:2012bt}. This situation (sometimes referred to as the
``Vacuum Catastrophe'' in the literature) requires extra hypotheses
about (partial or complete) compensation of vacuum condensates of
different types to the net vacuum energy density of the Universe
(see e.g. Ref.~\cite{Polchinski:2006gy}). Note that most of the
models developed in the literature adopt the same point of view
trying to cancel or suppress short distance vacuum fluctuations in
one way or another (for a review on such models, see e.g.
Ref.~\cite{Copeland:2006wr} and references therein). Still, a
dynamical mechanism for such gross cancellations and corresponding
major fine-tuning of vacuum parameters is not known at the moment
and is a subject of intensive studies in the vast literature (see
e.g. Ref.~\cite{Dolgov}).

Historically, about 45 years ago Ya.~B.~Zeldovich has pointed out in
Ref.~\cite{Zeldovich:1967gd} that cosmological constraints on the
value of the $\Lambda$-term density to a good accuracy correspond to
a simple purely mass dimensional estimate $\varepsilon_{\Lambda}\sim
G m^6$ (in natural units $\hbar=c=1$), where $G=M_{Pl}^{-2}$ is the
gravitational constant and $m$ is some characteristic mass of light
elementary particles (or light hadrons known at that time). For the
first time, he proposed an interpretation of the $\Lambda$-term as
an effect of gravitational interactions of virtual particles in the
physical vacuum. Almost at the same time, A.~D.~Sakharov noticed in
Ref.~\cite{Sakharov:1967pk} that, indeed, extra terms describing an
effect of graviton exchanges between identical particles (e.g.
bosons in the ground state) should appear in the right hand side of
Einstein equations averaged over their quantum ensemble. Below we
refer to this approach as the Zeldovich-Sakharov (ZS) scenario.

One of the well-known representations of the Zeldovich relation
$\varepsilon_{\Lambda}\sim G m^6$ through the basic fundamental
constants, the minimal (typical hadron scale) and maximal (Planck
scale) scales of fundamental particle physics, has been proposed by
N.~S.~Kardashev in Ref.~\cite{Kardashev}. It has the following form
\begin{eqnarray}
 \varepsilon_{\Lambda}=\frac{m_\pi^6}{(2\pi)^4M^2_{Pl}}\simeq
 3.0\times 10^{-35}\; \text{MeV}^4\,, \label{Lambda-mpi}
\end{eqnarray}
where $m_\pi \simeq 138\;\text{MeV}$ is the pion mass \cite{PDG},
$M_{Pl}=G^{-1/2} \simeq 1.22\cdot 10^{22}\;\text{MeV}$ is the Planck
mass. It is worth to notice that the representation
(\ref{Lambda-mpi}) turns out to be numerically very close to the
most recent WMAP data, well within experimental error bars,
$\varepsilon_{\Lambda}^{\rm{exp}}=(3.0\pm 0.7)\times
10^{-35}\;\text{MeV}^4$ \cite{WMAP}. Such a remarkable numerical
coincidence of the simple Zeldovich-Kardashev (ZK) formula
(\ref{Lambda-mpi}) to the current cosmological observations seems to
be almost too good to be just an accident \cite{Schutzhold:2002sg}.
This situation yet remains puzzling and requires a closer look into
possible physical reasons for that. A deeper theoretical
understanding of those reasons is the major goal of our current
work.

It is well-known that the pion mass is an object of essentially
non-perturbative Quantum Chromodynamics (QCD) driven by properties
of the quark-gluon vacuum condensate \cite{Shifman:1978bx}, the
strongest non-perturbative vacuum sub-system at the minimal energy
scales known in particle physics. So the numerical coincidence of
the ZK result (\ref{Lambda-mpi}) to the observational data if not
accidental poses a question about a possible role of strong
non-perturbative vacuum fluctuations of quark and gluon fields in
the $\Lambda$-term generation in the framework of ZS scenario
described above \cite{Zeldovich:1967gd,Sakharov:1967pk}.

Among existing quantum field theory approaches describing the
observable $\Lambda$-term density value, the most promising and
successful ones are recently proposed by Klinkhamer and Volovik in
Ref.~\cite{Klinkhamer:2009nn} and by Urban and Zhitnitsky in
Ref.~\cite{Urban:2009yg}. These approaches attempt to interpret the
positive Dark Energy density as a small but non-vanishing effect of
gravitating non-perturbative QCD vacuum fluctuations in a
non-trivial background of expanding Universe. The first approach
\cite{Klinkhamer:2009nn} is based upon a generic ``$q$-theory''
operating with a conserved microscopic $q$ value, whose statics and
dynamics are studied at macroscopic scales and which can, in
principle, be identified with the gluon condensate in QCD. In this
case it has been shown explicitly that the gravitating
non-perturbative QCD vacuum fluctuations dynamically generate a
nonzero limiting value of the vacuum energy density in the
nonequilibrium context of the expanding Universe. The second
approach \cite{Urban:2009yg} focuses on dynamics of the ghost fields
in the low energy (chiral) QCD. In particular, it was shown that the
Veneziano ghost, being unphysical in the usual Minkowski QFT,
exhibits non-vanishing physical effect in the expanding universe
leading to a positive vacuum energy density with a time-dependent
equation of state. Both approaches arrive at similar
order-or-magnitude estimates for the $\Lambda$-term density in the
expanding Universe in terms of the maximal $M_{Pl}$ and minimal
$\Lambda_{QCD}$ scales of particle physics whose existence is
required by the unitarity condition of an underlined quantum theory
\cite{Urban:2009yg}. Similar QCD-based ideas of the Dark Energy
origin from the effective interacting gluon condensate were
previously explored in Refs.~\cite{Zhang}. Also, quantum effects in
the Born-Infeld fields condensation very similar to the gluon
condensation in QCD (realized by means of the trace anomaly) which
could potentially play an important role in the dynamical
$\Lambda$-term generation were discussed in
Ref.~\cite{Elizalde:2003ku}. Another interesting claim was made in
Ref.~\cite{Poplawski}, where it was demonstrated that an extra
contribution to the vacuum energy density of the form $\sim G
\Lambda_{\rm QCD}^6$ may also originate from a consideration of QCD
in spacetime with torsion in the framework of the Einstein-Cartan
theory with minimal fermion coupling to torsion. Such promising
QCD-based interpretations of the Dark Energy give us a strong
motivation for further investigations in this direction and requires
development of a rigorous self-consistent theoretical framework
based upon quasiclassical (semiquantum) gravity and theory of
non-perturbative QCD vacuum fluctuations, in their existing standard
formulation.

The basic ideas and contents of this paper can be condensed into a
few short paragraphs as follows. In the zeroth order in metric
fluctuations, the contribution of non-perturbative QCD vacuum
fluctuations to the net energy density of the Universe can be, in
principle, {\it compensated} via a {\it macroscopic}
spatially-homogeneous modes of the cosmological Yang-Mills fields
discussed by us recently in Ref.~\cite{Pasechnik} or by a {\it
microscopic} dynamical mechanism of the QCD vacuum self-tuning. The
latter possibility will be considered below in Section
II\footnote{Possible mechanisms for compensation of other {\it
perturbative} contributions to the net vacuum energy density of the
Universe from virtual fermions and bosons (e.g. Higgs condensates,
graviton condensate etc) typically refer to Supersymmetry and
high-scale Grand Unified Theories
\cite{Copeland:2006wr,Nobbenhuis:2006yf} and will not be discussed
further in this paper.}. This ``compensation hypothesis'' is our
basic phenomenologically motivated assumption which is necessary to
remove the huge contributions from perturbative and non-perturbative
vacua into the vacuum energy density at macroscopic length scales
$l\gg \Lambda_{\rm QCD}^{-1}$ without an account for gravity-induced
effects. It will also be relevant for quantitative predictions of an
uncompensated gravity-induced residue and its identification with
the observable $\Lambda$-term.

Further, in Section III and IV we demonstrate that the
non-perturbative vacuum fluctuations of quark and gluon fields {\it
dynamically} induce the fluctuations of the background metric in the
expanding Universe through a coupling of gravity to the gluon field
via the {\it trace anomaly}. Such a coupling leads to an extra
correction term in the energy-momentum tensor which is {\it linear}
in the graviton field (as the first-order correction in the
quasiclassical gravity approximation). We make the corresponding
theoretical prediction and perform basic studies of properties of
this extra dynamical contribution to the energy-density of the
Universe from graviton-induced interactions in the quark-gluon
vacuum. This prediction is the main result of the paper which is
obtained in a theoretically rigorous way without imposing any extra
physical assumptions. Qualitatively, we observe that the first-order
gravity-induced contribution averaged over the quantum ensemble of
gluon field fluctuations at large scales does not depend on time
which is in accordance with the cosmological observations.

Further, we considered a possible identification of the extra
gravity-induced contribution to the vacuum energy density with the
time-independent $\Lambda$-term density. We have shown that this
identification is possible after the QCD phase transition in the
Universe evolution if the ``compensation hypothesis'' holds true.
The existing (static) instanton theory of the non-perturbative QCD
vacuum \cite{Belavin:1975fg} (for a review on the QCD theory of
instantons see e.g. Ref.~\cite{instantons}), however, does not
enable us to constrain the precise numerical value of the QCD
contribution to the $\Lambda$-term density fully dynamically, based
on the First Principles only. Nevertheless, it is possible, in fact,
to find some specific conditions and theoretical constraints under
which the QCD vacuum theory leads to the observed $\Lambda$-term
density value. Two of these conditions inherent to QCD, the presence
of the conformal anomaly and a strong coupling theory, have already
been emphasized e.g. in
Refs.~\cite{Schutzhold:2002sg,Thomas:2009uh}. In order to estimate
the numerical value of the $\Lambda$-term we assume that the third
condition should hold as well: {\it a shift between the
characteristic scales of quantum-topological and quantum-wave
fluctuations in the QCD vacuum is induced by the chiral symmetry
breaking}. This assumption is yet to be rigorously proven in the
framework of chiral QCD theory and lattice simulations but can be
taken as a good approximation for our quantitative analysis. As will
be shown in Section V and VI, under the latter assumption the extra
quantum gravity correction term indeed gives rise to an additional
positive non-vanishing $\Lambda$-term-type contribution to the
vacuum energy density of the Universe close to the observable value,
in accordance with the Sakharov-Zeldovich scenario
\cite{Zeldovich:1967gd,Sakharov:1967pk}.

\section{QCD vacuum energy: zeroth order in metric perturbations}
\label{sec:top_vs_col}

The ground state in QCD is typically characterized by non-vanishing
condensates of strongly interacting quarks and gluons commonly
referred to as {\it the quark-gluon condensate} responsible for the
confined phase of quark matter. In the framework of the instanton
liquid models, the {\it topological (or instanton) modes} of the
quark-gluon condensate are given essentially by the strong
non-perturbative fluctuations of the gluon and sea (mostly, light)
quark fields which are induced in processes of quantum tunneling of
the gluon vacuum between topologically different classical states
\cite{instantons}. The topological instanton-type contribution
$\varepsilon_{vac(top)}$ to the energy density of the QCD vacuum is
one of its main characteristics \cite{Shifman:1978bx} and can be
derived from the well-known trace anomaly relation
\cite{trace-anomaly}
\begin{equation}\label{traceA}
T_{i({\rm
QCD})}^i=\frac{\beta(g_s^2)}{2}F_{ik}^aF^{ik}_a+\sum_{q=u,d,s}m_q\bar{q}q\,,
\end{equation}
where $T_{i({\rm QCD})}^i$ is the trace of the energy-momentum
tensor, $\alpha_s=g_s^2/4\pi$, $m_q$ are the light current sea quark
masses, and $F_{ik}^a$ is the gluon field strength tensor in the
standard normalisation. A vacuum average of the trace $\langle
0|T_{i({\rm QCD})}^i|0\rangle=4\varepsilon_{vac(top)}$ finally leads
to the well-known formula for the instanton energy density (see e.g.
Refs.~\cite{Voloshin,Gogohia:1999ip})
\begin{eqnarray}
\varepsilon_{vac(top)}&=&-\frac{9}{32}\langle0|:\frac{\alpha_s}{\pi}F^a_{ik}(x)
F_a^{ik}(x):|0\rangle + \frac14 \Big[\langle0|:m_u\bar
uu:|0\rangle+\langle0|:m_d\bar dd:|0\rangle \nonumber \\&+&
\langle0|:m_s\bar ss:|0\rangle\Big] \simeq -(5\pm 1)\times 10^{9}\;
\text{MeV}^4\,,
 \label{top}
\end{eqnarray}
composed of gluon and light sea $u,d,s$ quark contributions. This is
the saturated (maximal) value of the topological contribution to the
QCD vacuum energy density, while possible refinements to it are
dependent on poorly known non-perturbative long-range Yang-Mills
dynamics and were discussed e.g. in Ref.~\cite{Gogohia:1999ip}.

Clearly, contributions of a different physical nature should
compensate the QCD contribution (\ref{top}) to the vacuum energy of
the Universe since its value by far is not compatible with the
cosmological observations and data on the $\Lambda$-term density
value \cite{WMAP}. On the other hand, within the general problem of
vacuum condensates cancellation and corresponding fine tuning of
vacuum substructures, the contribution (\ref{top}) has a special
status. Various existing cancellation mechanisms refer essentially
to an unknown high-scale physics beyond the Standard Model e.g. to
Supersymmetry \cite{Copeland:2006wr,Nobbenhuis:2006yf}. However,
they cannot be applied for a compensation of the specifically
non-perturbative and low-energy QCD contribution given by
Eq.~(\ref{top}). This issue forces us to look for alternative
cancellation mechanisms.

One of the possible ways to eliminate the {\it microscopic} QCD
vacuum contribution (\ref{top}) to the vacuum energy density of the
Universe was discussed earlier by us in Ref.~\cite{Pasechnik}
introducing the hypothesis about the existence of the cosmological
{\it macroscopic} Yang-Mills fields in early Universe. In
particular, it was claimed that the negative QCD contribution
(\ref{top}) can, in principle, be canceled by a positive constant
contribution generated by spatially-homogeneous modes of such a
field. Corresponding quasiclassical solution for these modes is
exact, it necessarily takes into account interactions with the QCD
vacuum (vacuum polarisation effects) and gives rise to the
spatially-homogeneous finite-time instantons.

In this paper, we consider another approach to the compensation of
the topological QCD contribution assuming that there might be extra
contributions from the {\it spatially-inhomogeneous} modes of a
different nature to the QCD vacuum density at the QCD energy scale
$\sim \Lambda_{\rm QCD}$ besides the topological (instanton) ones
given by Eq.~(\ref{top}) and spatially-homogeneous ones predicted in
Ref.~\cite{Pasechnik}. In fact, the importance of such extra {\it
long-range quantum-wave} fluctuations of the {\it hadronic vacuum}
has been emphasized earlier, e.g. in Ref.~\cite{Dorokhov}. Within
the alternative {\it microscopic} approach to compensation of the
QCD contribution, the total QCD vacuum energy density may turn into
zero due to a fine-tuning of the QCD vacuum parameters corresponding
to quantum-topological and quantum-wave fluctuations in a vicinity
of the $\Lambda_{\rm QCD}$ scale. Let us explore such a possibility
in detail.

The quantum-topological fluctuations contributing to the QCD vacuum
energy density (\ref{top}) exist at typical space-time scales $\sim
l_g$ which satisfy the following approximate inequality
\cite{instantons}
\begin{equation}
\begin{array}{c}
\displaystyle    l_{g(min)}\lesssim l_g < l_{g(max)},
\\[3mm]
\displaystyle l_{g(min)}\simeq(1500 \; \text{MeV})^{-1},\qquad
l_{g(max)}\simeq (500 \; \text{MeV})^{-1}\,. \label{sc}
\end{array}
\end{equation}
Here, the values of $l_{g(min)}, \; l_{g(max)}$, which can be
estimated e.g. in the lattice QCD framework, are interpreted as the
minimal and maximal length scales of the non-perturbative gluon
field fluctuations and approximately correspond to boundaries of the
light resonances region in the hadron spectrum
\cite{Shifman:1978bx}\footnote{The numerical values for boundaries
in Eq.~(\ref{sc}) can be somewhat model-dependent which, in
practice, would not affect any of our conclusions here.}.

In general, besides topological (instanton-type) fluctuations
contributing to Eq.~(\ref{top}), there are other two types of
quantum-wave QCD vacuum fluctuations: (1) {\it perturbative}
fluctuations of gluon and quark fields at smaller length scales $l <
l_{g(min)}$, and (2) vacuum fluctuations corresponding to {\it
collective wave motion} of gluon and quark fields at the same scales
as the instanton ones (\ref{sc}) \cite{Dorokhov}. As was mentioned
earlier, the problem of compensation of the short-range perturbative
QCD fluctuations, along with all other high energy vacuum subsystems
corresponding to e.g. zero-point fluctuations of fundamental fields
and Higgs-type condensates, is the subject of a supersymmetric
``Theory of Everything'' and therefore is not discussed here.
Meanwhile, quantum-wave fluctuations of the second type (from now
on, we denote them as the {\it collective} ones) have quantum
numbers of hadrons with masses $m_h \leq l_{g(min)}^{-1}$. Their
{\it renormalized} Lorentz-invariant contribution to the net QCD
vacuum energy density is then expressed in terms of the light hadron
masses and the universal cut-off parameter $\mu \simeq
l_{g(min)}^{-1}$ (playing a role of the renormalisation scale) as
follows \cite{Shifman:1978bx}
\begin{equation}
\begin{array}{c}
\displaystyle \varepsilon_{vac(h)}=\frac{1}{32\pi^2}\left(2\sum_B
(2J_B+1)m_B^4\ln\frac{\mu}{m_B}
-\sum_M
(2J_M+1)m_M^4\ln\frac{\mu}{m_M}\right).
\end{array}
\label{h}
\end{equation}
where $J_B$ and $J_M$ ($m_B$ and $m_M$) are the spins (masses) of
respective baryon and meson degrees of freedom, respectively.

Note, both topological and collective fluctuations have a
non-perturbative nature. Strictly speaking, the expression
(\ref{top}) based upon the trace anomaly relation (\ref{traceA})
should contain both the quantum-topological and collective
contributions. Their separate consideration, however, is necessary
since they have a completely different structure \cite{Dorokhov}
while there is no a self-consistent theory of the non-perturbative
QCD vacuum dynamics which could enable us to extract the
collective-wave contribution (\ref{h}) from the general expression
(\ref{top}). To this end, we formally remove the quantum-wave
fluctuations from Eq.~(\ref{top}) taking the saturated (maximal)
value for the topological contribution as was formally denoted by
the symbols of normal ordering.

The most important observation here is that taking into account only
metastable hadronic degrees of freedom in Eq.~(\ref{h}) -- the
baryon octet $B=\{N,\,\Lambda,\,\Sigma,\,\Xi\}$ and the pseudoscalar
nonet $M=\{\pi,\,K,\,\eta,\,\eta'\}$ -- we obtain a meaningful
result, namely, $\varepsilon_{vac(top)}+\varepsilon_{vac(h)}=0$ for
a reasonable cut-off parameter value $\mu\simeq 1.2 \;\text{GeV}$.
It turns out that the topological and collective quantum-wave
fluctuations contribute to the QCD vacuum energy density with {\it
opposite} signs. Therefore, a particular matching of numerical
values of the non-perturbative QCD parameters (e.g. light hadron
masses and scale parameters) via yet unknown dynamical mechanism
could, in principle, provide zeroth net value of the
non-perturbative QCD vacuum energy density without incorporating any
extra physics at different space-time scales.

Note, the hypothesis about the exact cancellation of
quantum-topological and quantum-wave contributions to the vacuum
energy as an internal feature of the theory of non-perturbative QCD
vacuum does not mean that the sum of their quantum fluctuations is
identically equal to zero. Indeed, let us consider the complete
unordered two-point function
\begin{equation}
\begin{array}{c}
\displaystyle
\langle0|\frac{\alpha_s}{\pi}F^a_{ik}(x)F_a^{ik}(x')|0\rangle\equiv
\langle0|:\frac{\alpha_s}{\pi}F^a_{ik}(x)F_a^{ik}(x'):|0\rangle+
\langle0|\frac{\alpha_s}{\pi}F^a_{ik}(x)F_a^{ik}(x')|0\rangle_{(h)}\,.
\end{array}
\label{xx'}
\end{equation}
The first term in Eq.~(\ref{xx'}) is expressed via the
experimentally measured value (introduced for the first time in
Ref.~\cite{Shifman:1978bx}) and a correlation function $D(x)$ which
can be constrained e.g. in lattice QCD or effective field theory
methods, i.e. \cite{Dosch}
\begin{eqnarray} \nonumber
&&\langle0|:\frac{\alpha_s}{\pi}F^a_{ik}(x)F_a^{ik}(x'):|0\rangle=
\langle0|:\frac{\alpha_s}{\pi}F^a_{ik}(0)F_a^{ik}(0):|0\rangle
D_{top}(x-x')\,,\quad D_{top}(0)=1\,,\\
&&\langle0|:\frac{\alpha_s}{\pi}F^a_{ik}(0)F_a^{0}(0):|0\rangle=(360\pm
20 \;\text{MeV})^4\,. \label{top1}
\end{eqnarray}
For a detailed overview of the methods of calculation of higher
power corrections to non-local condensates in QCD, see e.g.
Ref.~\cite{Grozin}. The second term in Eq.~(\ref{xx'}) representing
the quantum-wave component of the two-point function has an
analogical representation. Now, the hypothesis about the exact
cancellation of topological and quantum-wave contributions written
in terms of normally ordered correlation functions
\[
\displaystyle
\langle0|:\frac{\alpha_s}{\pi}F^a_{ik}(0)F_a^{ik}(0):|0\rangle=-
\langle0|\frac{\alpha_s}{\pi}F^a_{ik}(0)F_a^{ik}(0)|0\rangle_{(h)}
\]
leads to the following expression for the sum of corresponding
fluctuations:
\begin{equation}
\begin{array}{c}
\displaystyle
\langle0|\frac{\alpha_s}{\pi}F^a_{ik}(x)F_a^{ik}(x')|0\rangle=
\langle0|:\frac{\alpha_s}{\pi}F^a_{ik}(0)F_a^{ik}(0):|0\rangle
D(x-x')\,,
\\[3mm]
 \displaystyle D(x-x')=D_{top}(x-x')-D_{h}(x-x')\,. \qquad D(0)=0.
\end{array}
\label{d}
\end{equation}
As was mentioned earlier, the fluctuations of both types occur at
the same space-time scale. However, identical vanishing of the
function (\ref{d}) can not be assumed here, i.e. a completely
different space-time dynamics of the quantum-topological and
quantum-wave fluctuations, of course, does not imply that
$D(x-x')\equiv 0$ identically.

Thus, we have illustrated another possibility to eliminate the
non-perturbative QCD contribution to the ground state energy of the
Universe based on the hypothesis about a specific fine-tuning of the
QCD vacuum parameters. Both macroscopic previously discussed in
Ref.~\cite{Pasechnik} and microscopic mechanisms considered above in
this Section have a non-perturbative nature imprinted in essentially
unknown quantum dynamics of the QCD vacuum, and the real situation
can be, in principle, a superposition of both cancellation
mechanisms (not excluding other possibilities, of course).

Once the QCD vacuum energy contribution has been eliminated to the
leading order in metric perturbations, the observable small
$\Lambda$-term density can be further generated by quantum-gravity
corrections according to the SZ scenario
\cite{Zeldovich:1967gd,Sakharov:1967pk}, i.e. by quantum metric
fluctuations dynamically induced by the non-zeroth non-perturbative
gluon field fluctuations described by Eq.~(\ref{d}). We will further
demonstrate this fact by using the quasiclassical approximation
methods in the conventional General Relativity and quantum gravity
frameworks.

\section{QCD vacuum energy: first order in metric perturbations}

\subsection{Equations of motion for metric fluctuations and macroscopic geometry}

In this Section, we briefly overview the quasiclassical
(semiquantum) gravity framework in four dimensions, where the
Zeldovich-Sakharov scenario \cite{Zeldovich:1967gd,Sakharov:1967pk}
can be mathematically realized in the simplest and well-grounded
way.

The semiclassical approach to quantum gravity deals with quantum
fields defined on a classical background \cite{QG}. Typically, one
starts from the action of the gravitational and external physical
fields written in terms of the corresponding quantum field operators
as follows
\begin{equation}\label{gravS}
  S = \int{Ld^4x}\,, \hspace{5mm} L=-\frac{1}{2\varkappa}
  \sqrt{-\hat{g}}\hat{g}^{ik}\hat{R}_{ik}
  +L\left(\hat{g}^{ik},\chi_{A}\right)\,,
\end{equation}
where $\hat{g}^{ik}$ and $\hat{R}_{ik}$ are the metric and curvature
operators, respectively; $L\left(\hat{g}^{ik},\chi_{A}\right)$ is
the Lagrangian density of external physical (e.g. gauge) fields
$\chi_{A}$ with spins $J<2$ interacting with each other and with
gravity; index $A$ numerates degrees of freedom of these external
fields.

In the framework of quasiclassical gravity theory it is
conventionally assumed \cite{QG} that the metric operator
$\hat{g}^{ik}$ contains $c$-number part $g^{ik}$ -- the macroscopic
space-time metric, and operator part $\Phi_i^k$ -- the quantum
graviton field. In what follows, we work in the Heisenberg
representation. This means that one introduces an extra postulate
about the existence of the Heisenberg state vector $|0\rangle$,
which contains information about initial states of all incident
quantum fields. Then, the graviton field operator $\Phi_i^k$, by
definition, satisfies the following condition on its average over
the Heisenberg states:
\begin{equation}\label{Phiav}
\langle 0|\Phi_i^k|0\rangle = 0\,.
\end{equation}

As usual, derivation of the quasiclassical gravity theory equations
of motion includes variation and averaging operations. In order to
match the classical evolution of the macroscopic metric with quantum
dynamics of gravitons it is necessary to require that independent
variations of the action (\ref{gravS}) over $g^{ik}$ (taken at
$\Phi_i^k = \text{const}$) and $\Phi_i^k$ (taken at $g^{ik} =
\text{const}$) must lead to the same operator equations, namely,
\begin{equation}
\begin{array}{c}
\displaystyle  \delta \int{Ld^4x} = -\frac{1}{2}\int{d^4 x
  \left(\sqrt{-g}\delta g^{ik}\hat{G}_{ik}\right)_{\Phi_i^k=\text{const}}}
  =
 -\frac{1}{2}\int{d^4 x}
  \left(\sqrt{-g}\delta
  \Phi^{ik}\hat{G}_{ik}\right)_{g^{ik}=\text{const}}=0\,,
\end{array}
\label{S}
\end{equation}
giving rise to
\begin{equation}\label{hGik}
\begin{array}{c}
\displaystyle
  \hat{G}_i^k=\frac{1}{2}
  \left(\delta_l^k\delta_i^m+g^{km}g_{il}\right)\left(\frac{\hat{g}}{g}\right)^{1/2}\hat{E}_m^l=0\,,
\\[3mm]
\displaystyle \hat{E}_m^l=
    \frac{1}{\varkappa}\left(\hat{g}^{lp}\hat{R}_{pm}
  -\frac{1}{2}\delta_m^l\hat{g}^{pq}\hat{R}_{pq}\right)
  -\hat{g}^{lp}\hat{T}_{pm}\left(\hat{g}^{ik},\chi_{A}\right)\,,
\end{array}
\end{equation}
where $T_{pm}\left(\hat{g}^{ik},\chi_{A}\right)$ is the operator
analog of the classical energy-momentum tensor corresponding to the
Lagrangian $L\left(\hat{g}^{ik},\chi_{A}\right)$. By averaging the
operator equations (\ref{hGik}) over the Heisenberg states, one
obtains the equations of motion for the macroscopic (background)
metric $g^{ik}$:
\begin{equation}\label{av}
\langle 0|\hat{G}_i^k|0\rangle = 0\,.
\end{equation}
Subtracting $c$-number part (\ref{av}) from operator (\ref{hGik}),
one obtains the equations of motion for the graviton fields
$\Phi_i^k$:
\begin{equation}\label{graveq}
  \hat{G}_i^k - \langle 0|\hat{G}_i^k|0\rangle = 0\,.
\end{equation}
In addition to Eqs.~(\ref{av}) and (\ref{graveq}), one also gets the
operator equations of motion for external fields with $J<2$ by means
of variations of the action (\ref{gravS}) over $\chi_{A}$.

An explicit form of the functional $\hat{G}_i^k$ can be obtained by
a variation of the action over the macroscopic metric $g^{ik}$
without implying an explicit form for the quantum
$\hat{g}^{ik}\left(g^{lm},\;\Phi_n^s \right)$ operator. The
condition (\ref{S}) providing consistency of equations (\ref{av})
and (\ref{graveq}) fixes the exponential parameterisation for the
metric operator as follows \cite{Borisov}
\begin{equation}\label{exp}
\begin{array}{c}
  \sqrt{-\hat{g}}\hat{g}^{ik} = \sqrt{-g}g^{il}(\exp{\psi})_l^k=
  \sqrt{-g}g^{il}
  \left(
         \delta_l^k + \psi_l^k + \frac{1}{2}\psi_l^m\psi_m^k+\ldots
  \right)\,,
\end{array}
\end{equation}
where a shorthand notation for the graviton field has been
introduced
$$
 \psi_i^k = \Phi_i^k - \frac{1}{2}\delta_i^k \Phi\,.
$$
Applying the exponential parameterization (\ref{exp}), operator
$\hat{G}_i^k$ can then be transformed to the following compact
expression:
\begin{eqnarray}\nonumber
\hat{G}_i^k &=& \frac{1}{2\varkappa}
  \left(
         \psi_{i;l}^{k;l}
         -\psi_{i;l}^{l;k}
         -\psi_{l;i}^{k;l}
         +\delta_i^k \psi_{l;m}^{m;l}+\psi_i^l R_l^k
         +\psi_l^k R_i^l
         -\delta_i^k\psi_l^m R_m^l
  \right) \\ &+&\frac{1}{\varkappa}\left(
  R_i^k -\frac{1}{2}\delta_i^k R
  \right)
  - \hat{T}_i^k\,, \label{Gpsi}
\end{eqnarray}
where $R_i^k$ is the Ricci tensor of macroscopic space-time with the
background metric $g_{ik}$ in which all the covariant derivatives
and lowering/raising index operations are defined; and
\begin{equation}\label{Tandgrav}
  \hat{T}_i^k=\hat{T}_{i(G)}^k+\frac{1}{2}
  \left(\delta_l^k\delta_i^m+g^{km}g_{il}\right)
  \left(\frac{\hat{g}}{g}\right)^{1/2}
  \hat{g}^{lp}\hat{T}_{pm}\left(\hat{g}^{ik},\chi_{A}\right)
\end{equation}
is the total energy-momentum tensor operator incorporating the
graviton field contribution
\begin{equation}\label{Tgrav}
\begin{array}{c}
\displaystyle
  \hat{T}_{i(G)}^k = \frac{1}{4\varkappa}
  \left(
        \psi_{m;i}^l\psi_l^{m;k}
        -\frac{1}{2}\psi_{;i}\psi^{;k}
        -\psi_{i;m}^l\psi_l^{m;k}  -\psi_l^{k;m}\psi_{m;i}^l\right)
\\[3mm] \displaystyle
  -\frac{1}{8\varkappa}\delta_i^k
  \left(
        \psi_{m;n}^l\psi_l^{m;n}
        -\frac{1}{2}\psi_{;n}\psi^{;n}
        -2\psi^l_{n;m}\psi_l^{m;n}
  \right)
\\[3mm] \displaystyle
  -\frac{1}{4\varkappa}
  \left(
        2\psi_n^l\psi_i^{k;n}
        -\psi_n^k\psi_i^{l;n}
        -\psi_i^n\psi_{\hspace{2mm};n}^{kl}
         +\psi_i^{n;k}\psi_n^l
        +\psi_{n;i}^{k}\psi^{nl}+\delta_i^k\left(\psi_m^n\psi_n^l\right)^{;m} \right)_{;l}
\\[3mm] \displaystyle
-\frac{1}{4\varkappa}
  \left(
        \psi_i^m\psi_n^lR_l^k
        +\psi_n^k\psi_l^nR_i^l
        -\delta_i^k\psi_l^n\psi_n^mR_m^l
  \right)
  +O(\psi^3)\,.
\end{array}
\end{equation}
Averaging the operator $\hat{G}_i^k$ given by Eq.~(\ref{Gpsi}) with
an extra defining condition on the quantum graviton field $\Phi_i^k$
(\ref{Phiav}), we see that the equations of motion for the
macroscopic (background) metric $g_{ik}$ (\ref{av}) are transformed
into usual Einstein equations as expected
\begin{equation}
\displaystyle  \frac{1}{\varkappa}\left(
  R_i^k -\frac{1}{2}\delta_i^k R
  \right)=\langle 0|\hat{T}_i^k|0\rangle\,,
\label{E}
\end{equation}
where a macroscopic average $\langle 0|\hat{T}_i^k|0\rangle$ in the
right hand side contains a contribution from the energy-momentum
tensor of the quantum graviton field $\hat{T}_{i(G)}^k$ according to
Eq.~(\ref{Tandgrav}). Applying the same procedure to
Eq.~(\ref{graveq}) using expression (\ref{Gpsi}), we finally obtain
the equations of motion for graviton fields in explicit form:
\begin{equation}\label{graveq2}
\begin{array}{c}
         \psi_{i;l}^{k;l}
         -\psi_{i;l}^{l;k}
         -\psi_{l;i}^{k;l}
         +\delta_i^k\psi_{l;m}^{m;l}
         +\psi_i^lR_l^k
         +\psi_l^kR_i^l
         -\delta_i^kR_l^m\psi_m^l
=2\varkappa\left(\hat{T}_i^k-\langle
0|\hat{T}_i^k|0\rangle\right)\,.
\end{array}
\end{equation}

A more rigorous derivation of the equations of motion (\ref{E}) and
(\ref{graveq2}) is based on the canonical quantum gravity in the
path integral (Faddeev-Popov) formulation \cite{FP}. In order to
turn from the complete quantum gravity to its quasiclassical
(semiquantum) limit, one starts from factorization of the path
integral measure provided by the exponential parametrization of the
metric operator (\ref{exp}). A subsequent calculation of the
factorized path integral, which is exact over the high-frequency
(quantum) fields $\Phi_i^k$ and saddle-point approximated over the
slow-changing (classical) fields $g_{ik}$, is equivalent to solving
the equations of motion in the operator formulation (\ref{E}) and
(\ref{graveq2}). This demonstrates theoretical consistency of the
operator approach described above.

\subsection{Operator gluodynamics with vacuum anomaly}

The non-perturbative fluctuations of gluon and quark fields
naturally gravitate and should be included into equations of motion
in the framework of operator field dynamics (\ref{E}) and
(\ref{graveq2}). The non-perturbative dynamics of gluon and quark
fields is not sufficiently well developed in the literature, so we
build up our analysis based upon the functional relations between
metric fluctuations and vacuum fluctuations of quark and gluon
fields only. These relations can be obtained from
Eq.~(\ref{graveq2}) and imply the existence of an adequate
field-theoretical model for the QCD vacuum energy-momentum tensor
incorporating conformal anomalies. The recipe for getting such a
tensor is demonstrated e.g. in Ref.~\cite{Voloshin}: within the
variational procedure for getting the energy-momentum tensor and
equations of motion, the QCD coupling $g_s$ can be viewed as an
operator depending on operators of quantum fields according to the
Renormalisation Group (RG) equations. In the framework of this
formalism, one introduces the gluon field (vector-potential)
operator $\mathcal{A}_i^a$, as a variational variable, related to
the gluon field operator in the standard normalisation as follows
$\mathcal{A}_i^a=g_sA_i^a$. The stress tensor operator is then
defined as
$\mathcal{F}^a_{ik}=\partial_i\mathcal{A}^a_k-\partial_k\mathcal{A}^a_i
+f^{abc}\mathcal{A}_i^b\mathcal{A}_k^c$. In the case of pure
gluodynamics, the QCD coupling operator $g_s^2=g_s^2(J)$ depends
upon the gauge operator of least dimension
$J\equiv\mathcal{F}^a_{ik}\mathcal{F}_a^{ik}$ by means of the
operator RG evolution equation
\begin{equation}
\displaystyle 2J\frac{dg_s^2(J)}{dJ}=g_s^2(J)\beta[g_s^2(J)]\,,
\label{rg}
\end{equation}
where $\beta[g_s^2(J)]$ is the QCD $\beta$-function calculable in
the standard quantum field theory framework. A solution of
Eq.~(\ref{rg}) is the substituted into the effective operator
Lagrangian
\begin{equation}
\displaystyle L_{\rm
eff}=-\frac{1}{4g_s^2(J)}\,\mathcal{F}^a_{ik}\mathcal{F}_a^{ik}\,,
\label{Lrg}
\end{equation}
whose variation w.r.t $\mathcal{A}_i^a$ leads to the operator
energy-momentum tensor of the gluon field
\begin{equation}
\displaystyle
\hat{T}_{i(g)}^k=\frac{1}{g_s^2(J)}\left(-\mathcal{F}^a_{il}\mathcal{F}_a^{kl}
+\frac14\delta_i^k
\mathcal{F}^a_{ml}\mathcal{F}_a^{ml}+\frac{\beta[g_s^2(J)]}{2}
\mathcal{F}^a_{il}\mathcal{F}_a^{kl}\right)
\label{S-tem}
\end{equation}
and operator equation of motion of pure gluodynamics
\begin{equation}
\begin{array}{c}
\displaystyle
D^{ab}_k\left\{g_s^{-2}(J)\left(1-\frac{\beta[g_s^2(J)]}{2}\right)\mathcal{F}^{ik}_b\right\}=0,
\\[3mm]
\displaystyle
D^{ab}_k=\delta^{ab}\partial_k-f^{abc}\mathcal{A}_k^c\,.
\end{array}
\label{S-eq}
\end{equation}
After a straightforward covariant generalization, the operator
gluodynamics defined by Eqs.~(\ref{S-tem}) and (\ref{S-eq}) can be
incorporated into the quasiclassical (semiquantum) gravity in the
Heisenberg operator formulation given by Eqs.~(\ref{E}) and
(\ref{graveq2}). In the framework of this formulation it is crucial
that the energy-momentum tensor (\ref{S-tem}) is conservative under
the operator equations of motion (\ref{S-eq}).

In the one-loop approximation we have
\begin{equation}
\displaystyle \beta[g_s^2(J)]=-\frac{bg_s^2(J)}{16\pi^2}\,, \qquad
\frac{g_s^2(J)}{4\pi}\equiv \alpha_s(J)=\frac{8\pi}{\displaystyle
b\ln(J/\lambda^4)}\,, \label{be}
\end{equation}
where $\lambda$ is the QCD scale parameter discussed below;
$b=b(0)=11$ is the one-loop $\beta$-function coefficient of the pure
gluodynamics (without quark fields).

In order to construct a realistic operator energy-momentum tensor in
QCD one has to incorporate quark fields. Formally, inclusion of the
quark fields lead to extra terms in the energy-momentum tensor and
to new operator equations of motion. However, this procedure can be
simplified within the adopted phenomenological approach. It is clear
from the beginning that an account for quark fields changes the
numerical values of the $\beta$-function coefficients. Also,
non-perturbative quark-gluon fluctuations happen at a characteristic
scale of four-momentum transfers smaller than the double charm quark
mass, so one has to fix $b=b(3)=9$ in Eq.~(\ref{be}) and,
correspondingly, in Eqs.~(\ref{S-tem}) and (\ref{S-eq}). Further, an
induced character of quark fluctuations in the quark sea
(effectively arising mostly due to gluon splittings in $q\bar{q}$
pairs at small momentum transfers) provides that QCD vacuum
observables can be approximately expressed through the square of
averaged gluon field fluctuations. In particular, for
quantum-topological quark-gluon fluctuations it is well known that
\cite{instantons}
\begin{equation}
\begin{array}{c}
\displaystyle \langle 0|:\bar{s}s:|0\rangle \simeq \langle 0|:
\bar{u}u :|0\rangle= \langle 0|: \bar{d}d: |0\rangle =
 -\langle 0|:\frac{\alpha_s}{\pi} F_{ik}^a
F_a^{ik}:|0\rangle L_{g}=
   - (225 \pm 25\;\text{MeV} )^3,
\end{array}
\label{q}
\end{equation}
where $L_{g} \simeq (1500\pm 300 \;\text{MeV})^{-1}$ is the
correlation length of fluctuations which is normally calculated
through the experimental data on quark and gluon condensates (and
supported by the lattice QCD calculations). Its value is close to
the minimal scale of fluctuations given in Eq.~(\ref{sc}) at which
their level is maximal. Within phenomenologically reasonable
assumptions discussed above in Sect.~\ref{sec:top_vs_col}, the
non-perturbative quantum-wave (hadron) fluctuations occur at the
same space-time scales, thus they should satisfy to a functional
relation analogical to Eq.~(\ref{q}). Under these assumptions the
operator relation between quark and gluon fluctuations can be
established by taking the trace of the averaged quark
energy-momentum tensor and then applying its conservation condition
and relations (\ref{top}) and (\ref{q}) valid for both
quantum-topological and quantum-wave contributions. This procedure
provides us with the effective quark contribution to the QCD energy
momentum tensor in the following form:
\begin{equation}
\displaystyle \hat{T}^k_{i(q),{\rm
eff}}=\frac{8L_{g}}{b(3)}(m_u+m_d+m_s)\hat{T}^k_{i(g)}\,.
\label{q-gik}
\end{equation}
Finally, adding Eqs.~(\ref{S-tem}) and (\ref{q-gik}) written in the
one-loop approximation gives phenomenologically motivated complete
QCD energy-momentum tensor in operator form
\begin{equation}
\begin{array}{c}
 \displaystyle
\hat{T}_{i({\rm QCD})}^k\simeq\frac{b_{\rm
eff}}{32\pi^2}\left(-\mathcal{F}^a_{il}\mathcal{F}_a^{kl}+
\frac14\delta_i^k
\mathcal{F}^a_{ml}\mathcal{F}_a^{ml}\right)\ln\frac{eJ}{\lambda^4}-
\delta_i^k\frac{b_{\rm
eff}}{128\pi^2}\mathcal{F}^a_{ml}\mathcal{F}_a^{ml}\,,
\\[5mm]
\displaystyle b_{\rm eff}=b(3)+8L_{g}(m_u+m_d+m_s)\simeq 9.6\,,
\end{array}
\label{qcd-ex1}
\end{equation}
and the operator equation of motion
\begin{equation}
 \displaystyle
D^{ab}_k\left(\mathcal{F}^{ik}_b\ln\frac{eJ}{\lambda^4}\right)=0\,.
\label{qcd-ex2}
\end{equation}
Of course, the resulting model expressions (\ref{qcd-ex1}),
(\ref{qcd-ex2}) are approximate and restricted by quantitative
phenomenological estimates (\ref{top}) and (\ref{q}) which
characterize the non-perturbative quark-gluon fluctuations.

Further simplifications are possible and make use of series
expansion of the logarithmic operator function in small fluctuations
as follows
\[
\displaystyle
\ln\frac{eJ}{\lambda^4}=\ln\frac{e\langle0|J|0\rangle}{\lambda^4}
+\frac{J-\langle0|J|0\rangle}{\langle0|J|0\rangle}+... \;.
\]
Note, the logarithmic function in the energy-momentum tensor
(\ref{qcd-ex1}) comes as a multiplier to an expression which has
zeroth vacuum expectation value. Thereby, an account for the leading
correction to the logarithm means that in the energy-momentum
tensor, together with leading terms linear in $\delta
J=J-\langle0|J|0\rangle$, one also takes into account the
higher-order terms in $\delta J$. Since the color factor suppresses
the mean square fluctuation of the logarithm by a factor of $1/24$,
then replacement of the operator function under the logarithm by its
averaged value can considered as a good approximation. Under such a
replacement, $g_s^2(eJ)$ operator transforms into the usual QCD
coupling at a characteristic scale of non-perturbative QCD
fluctuations in the region of four-momentum transfers squared less
than $L_{g}^{-2}\simeq (1500 \;\text{MeV})^2$, i.e.
\[
\displaystyle
\ln\frac{e\langle0|J|0\rangle}{\lambda^4}\simeq4\ln\frac{L_{g}^{-1}}{\Lambda_{\rm
QCD}},
\]
where $\Lambda_{\rm QCD}\simeq 160 \;\text{MeV}$ is the QCD scale
parameter.

All subsequent calculations are controlled by the fact that the
approximate expressions derived above must satisfy the
energy-momentum tensor conservation. The replacement of the operator
function under the logarithm by its averaged value transforms
operator equations (\ref{qcd-ex2}) into the standard Yang-Mills
equations. To this approximation, the energy-momentum
(\ref{qcd-ex1}) should also transform (up to a constant
multiplicative term and a constant additive term) into the standard
Yang-Mills energy-momentum tensor. This result is achieved by a
replacement of the conformal anomaly operator in the energy-momentum
tensor (\ref{qcd-ex1}) by its averaged value. Then, turning back to
original symbols $A_i^a=\mathcal{A}_i^a/g_s,\;
F_{ik}^a=\mathcal{F}_{ik}^a/g_s$ under the approximations adopted
above, one writes finally
\begin{equation}
\begin{array}{c}
\displaystyle \hat{T}_{i({\rm QCD})}^k=\frac{b_{\rm
eff}\alpha_s}{2\pi}\left(-F^a_{il}F_a^{kl}+\frac14\delta_i^k
F^a_{ml}F_a^{ml}\right)\ln\frac{L_{g}^{-1}}{\Lambda_{\rm QCD}}
-\delta_i^k\frac{b_{\rm
eff}}{32}\langle0|\frac{\alpha_s}{\pi}F^a_{ml}F_a^{ml}|0\rangle\,.
\label{S-tem0}
\end{array}
\end{equation}
\begin{equation}
\displaystyle D^{ab}_kF^{ik}_b=0\,, \qquad
D^{ab}_k=\delta^{ab}\partial_k-g_sf^{abc}A_k^c\,. \label{qcd-0}
\end{equation}

As a matter of fact, we have not done anything radically new here --
similar equations are often used in the Euclidean QCD framework
where the vacuum energy-momentum tensor is estimated at instanton
solutions of the classical Yang-Mills equations \cite{instantons}. A
proper covariant generalization of this framework will be applied
for the $\Lambda$-term density calculation in the next Section.

\section{$\Lambda$-term calculation}

Now, we have prepared everything what is needed for estimation of
the dynamically induced $\Lambda$-term. In the right hand side of
the macroscopic Einstein equations (\ref{E}) the terms which
correspond to the graviton-mediated interactions of the
non-perturbative quark-gluon fluctuations are of the order of $O(G)$
and $O(\alpha_sG)$ with $G$ being the gravitational constant. Of
course, in order to estimate the leading order effect, it makes
sense to take into account only the first-order (linear)
non-vanishing terms in gravitational constant $G$. Furthermore, due
to an obvious smallness of the typical QCD space-time scales
compared to the cosmological scales, the induced quantum
fluctuations of metric should be considered at the Minkowski
background. At last, it is sufficient to consider only the trace of
the macroscopic Einstein equations
$R+4\varkappa\,\varepsilon_{\Lambda}=0$ giving rise to the
QCD-induced $\Lambda$-term density
\begin{equation}
\displaystyle  \varepsilon_{\Lambda}=-\frac{b_{\rm eff}}{32}
  \langle0|\frac{\alpha_s}{\pi}\left(\frac{\hat{g}}{g}\right)^{1/2}
\hat{g}^{il}\hat{g}^{km}\hat{F}^a_{ik}\hat{F}^a_{lm}|0\rangle+
\frac14\langle0|\hat{T}_{(G)}|0\rangle\,. \label{E1}
\end{equation}

Let us start with expansion of the gluon stress tensor in series
over the metric fluctuations (gravitons). From Eq.~(\ref{qcd-0})
written in Riemann space
\[
\displaystyle    \left(\delta^{ab}\frac{\partial}{\partial
x^k}-g_sf^{abc}\hat{A}_k^c\right)\sqrt{-\hat{g}}\hat{g}^{il}\hat{g}^{km}\hat{F}^b_{lm}=0\,,
\]
it follows immediately
\begin{equation} \displaystyle
\hat{F}^a_{ik}=F^a_{ik}+\frac12\psi
F^a_{ik}-\psi_i^lF^a_{lk}-\psi_k^lF^a_{il}+O(\alpha_sG)\,,
\label{Fh}
\end{equation}
where $F^a_{ik}$ is the usual stress tensor at the macroscopic
background which does not account for interactions between gluon
field and metric fluctuations (gravitons). The expansion (\ref{Fh})
can be used only for extraction of the leading-order effect in the
first term of Eq.~(\ref{E1}), which initially is of the order of
$O(\alpha_s)$. Thereby, higher terms $O(\alpha_sG)$ in the expansion
(\ref{Fh}) generate corrections to the $\Lambda$-term density
(\ref{E1}) of the order of $O(\alpha_s^2G)$, which go beyond the
one-loop approximation adopted here and therefore are omitted.

The second term of Eq.~(\ref{E1}) is formed by induced fluctuations
of the metric (i.e. by a solution of equation (\ref{graveq2})),
which are unambiguously related with the gluon field fluctuations.
To the leading order in gravitational constant $G$, in the left hand
side of Eq.~(\ref{graveq2}) we keep only proper fluctuations of the
quadratic form $F^a_{il}F_a^{kl}$ without taking into account
gravity. Then, using Eq.~(\ref{S-tem0}) in the right hand side of
Eq.~(\ref{graveq2}), we obtain an important relation between the
graviton field $\psi_{ik}$ and the gluon field strength $F^a_{ik}$
to the respective order:
\begin{equation}
\begin{array}{c}
 \displaystyle
         \psi_{i,\,l}^{k,\,l}
         -\psi_{i,\,l}^{l,\,k}
         -\psi_{l,\,i}^{k,\,l}
         +\delta_i^k\psi_{l,\,m}^{m,\,l}
  =\frac{\varkappa\alpha_s\, b_{\rm eff}}{\pi}\left(-F^a_{il}F_a^{kl}+\frac14\delta_i^k
F^a_{ml}F_a^{ml}\right)\ln\frac{L_g^{-1}}{\Lambda_{\rm QCD}}\,.
\label{fluc}
\end{array}
\end{equation}

At the next step one substitutes the expansion (\ref{Fh}) into the
first term of Eq.~(\ref{E1}). The second term in Eq.~(\ref{E1}) can
be calculated using the trace of the averaged energy-momentum tensor
of gravitons (\ref{Tgrav}) (the averaging removes total derivatives
due to symmetry of the Minkowski space-time background). After the
averaging, the zeroth-order term in metric fluctuations (given by
the unperturbed trace of the quark-gluon vacuum energy-momentum
tensor) disappears due to the hypothesis about exact cancellation of
quantum-topological and quantum-wave contributions discussed in
Sect.~\ref{sec:top_vs_col}. Implying Eq.~(\ref{fluc}), the resulting
non-vanishing effect in the $\Lambda$-term density turns out to be
quadratic in the graviton field $\psi_{ik}$:
\begin{eqnarray}\nonumber
\varepsilon_{\Lambda} = &-& \frac{b_{\rm eff}}{64}\langle 0|\left(
\frac{\alpha_s}{\pi} F^a_{ml}F_a^{ml}\psi-4\frac{\alpha_s}{\pi}
F^a_{nm}F_a^{lm}\psi^n_l\right)|0\rangle \\
&-&\frac{1}{16\varkappa}\langle 0|\left(
        \psi_{m,\;n}^l\psi_l^{m,\;n}
        -\frac{1}{2}\psi_{,\;n}\psi^{,\;n}
        -2\psi^l_{n,\;m}\psi_l^{m,\;n}\right)|0\rangle\,.
\label{laint}
\end{eqnarray}
The second term in Eq.~(\ref{laint}) can be identically transformed
and simplified as follows: by transferring derivatives to one of the
multipliers in each term one arrives at the differential form which
can then expressed through the left hand side of Eq.~(\ref{fluc}).
Then applying the latter and combining all the terms together, the
resulting $\Lambda$-term density takes a remarkably simple form
convenient for phenomenological analysis:
\begin{equation}
\displaystyle  \varepsilon_{\Lambda}= -\frac{b_{\rm
eff}}{16}\ln\frac{L_{g}^{-1}}{e\Lambda_{\rm QCD}}\,\langle 0|
\frac{\alpha_s}{\pi} F^a_{il}F_a^{kl}\left(\psi_k^i-
\frac{1}{4}\delta_k^i\psi\right)|0\rangle\,. \label{laint1}
\end{equation}

The simplest way to obtain the physically meaningful result is to
fix the Fock gauge $\psi_{i;\,k}^k=0$. In this case, according to
Eq.~(\ref{fluc}), $\psi=0$, so the final solution for the graviton
field reads
\begin{equation}
\begin{array}{c}
\displaystyle   \psi_i^k(x)=\varkappa b_{\rm
eff}\ln\frac{L_{g}^{-1}}{\Lambda_{\rm QCD}}\,\int d^4x'\mathcal{G}
(x-x')\cdot \left(\frac{\alpha_s}{\pi}F^a_{il}(x')F_a^{kl}(x')-
\delta_i^k\frac{\alpha_s}{4\pi}F^a_{ml}(x')F_a^{ml}(x')\right).
\end{array}
\label{psiF}
\end{equation}
where $\mathcal{G}(x-x')$ is the Green function satisfying the Green
equation $\mathcal{G}_{,\,l}^{,\,l}=-\delta(x-x')$. After
substitution of Eq.~(\ref{psiF}) into Eq.~(\ref{laint1}), the
respective averages should be calculated according to the following
rules:
\begin{equation}
\begin{array}{c}
\displaystyle  \langle 0|
\frac{\alpha_s}{\pi}F^a_{il}(x)F_a^{kl}(x)\cdot \left(
\frac{\alpha_s}{\pi}F_{km}^b(x')F^{im}_b(x') -\langle
0|\frac{\alpha_s}{\pi}
F_{km}^b(x')F^{im}_b(x')|0\rangle\right)|0\rangle\,=
\\[3mm]
\displaystyle  \langle 0|\frac{\alpha_s}{\pi}F_{il}^a(x)F_{km}^b(x')
|0\rangle \langle 0|\frac{\alpha_s}{\pi}F^{kl}_a(x)F^{im}_b(x')
|0\rangle\, +
\\[3mm]
\displaystyle \langle 0|\frac{\alpha_s}{\pi}
F_{il}^a(x)F^{im}_b(x')|0\rangle \langle 0|\frac{\alpha_s}{\pi}
F^{kl}_a(x)F_{km}^b(x')|0\rangle\,,
\end{array}
\label{calc1}
\end{equation}
\begin{equation}
\begin{array}{c}
\displaystyle \langle
0|\frac{\alpha_s}{\pi}F_{il}^a(x)F_{km}^b(x')|0\rangle =
\frac{\delta^{ab}}{96}(g_{ik}g_{lm}-g_{im}g_{kl}) \langle
0|:\frac{\alpha_s}{\pi} F_{nj}^c(0)F^{nj}_c(0):|0\rangle D(x-x')\,.
\label{calc2}
\end{array}
\end{equation}
The first rule, Eq.~(\ref{calc1}), demonstrates the exchange
character of gravitational interactions of the gluon fluctuations,
the first-order one in fluctuations (the zeroth-order single-point
averages are explicitly subtracted). Also, Eq.~(\ref{calc2}),
besides the gauge/Lorentz symmetry properties, incorporates the
effect of compensation of the quantum-topological and quantum-wave
contributions -- the two-point function $D(x-x')$ is defined as a
difference between the corresponding correlation functions as
reflected in Eq.~(\ref{d}). Substitution of these formulas into
Eq.~(\ref{laint1}) leads to our final result for the QCD-induced
dynamical $\Lambda$-term energy density:
\begin{equation}
\begin{array}{c}
\displaystyle \varepsilon_{\Lambda}=-\pi G\langle
0|:\frac{\alpha_s}{\pi}F_{ik}^aF_a^{ik}:|0\rangle^2\times
\left(\frac{b_{\rm eff}}{8}\right)^2
\ln\frac{L_{g}^{-1}}{e\Lambda_{\rm
QCD}}\ln\frac{L_{g}^{-1}}{\Lambda_{\rm QCD}} \int
d^4y\mathcal{G}(y)D^2(y)=
\\[5mm]
\displaystyle =(1\pm0.5)\times 10^{-29}\Delta \;\text{MeV}^4.
\end{array}
\label{fin}
\end{equation}
Here, we introduced the dimensionless parameter
\begin{equation}
\displaystyle \Delta = -\frac{1}{L_g^2}\int
d^4y\mathcal{G}(y)D^2(y)\,, \label{Del}
\end{equation}
which is defined in the Euclidean 4-space such that $\Delta>0$. Its
numerical value has to be established by a dynamics in a complete
non-perturbative QCD vacuum theory, or estimated in the static
approximation e.g. in lattice QCD or in effective field theory
approaches.

In general, the expression (\ref{fin}) provides a good estimate for
the dynamical contribution of the graviton-exchange interactions in
the hadronic vacuum to the vacuum energy density of the Universe, it
is unavoidable and should be taken into consideration in any
theoretical or phenomenological analysis of the Dark Energy.

\section{Discussion and conclusions}

The major result of our paper is summarized in the formula
(\ref{laint1}) for the first-order quasiclassical gravity correction
to the vacuum energy density dynamically induced by graviton
exchanges between the gluon field fluctuations in the QCD vacuum.
This result is based upon the effective one-loop energy-momentum
tensor of quark and gluon fields (\ref{S-tem0}) incorporating the
conformal anomaly in the standard way and the conventional
quasiclassical gravity framework. We stress that the basic
theoretical result of the paper (\ref{laint1}) is rigorous and does
not incorporate any strong {\it ad hoc} physical assumptions.

Further, we subsequently formulate a new scenario for the
cosmological $\Lambda$-term generation based on Eq.~(\ref{laint1})
which is realistic under two necessary and sufficient conditions:
\begin{itemize}
\item the {\it exact compensation} of the non-perturbative quantum-topological
and quantum-wave contributions to the QCD vacuum energy density in
the {\it zeroth order in quasiclassical gravity expansion} i.e.
without graviton-induced interactions;
\item the {\it almost exact compensation} of the graviton-exchange interactions of
topological and wave fluctuations with a {\it small residual
uncompensated effect} summarized in Eq.~(\ref{fin}) which can
identified with the observable time-independent $\Lambda$-term (the
{\it first order effect in quasiclassical gravity expansion}). The
latter effect can be quantitatively estimated based on theoretical
and phenomenological constraints from the chiral QCD theory and
lattice QCD.
\end{itemize}

The ``Vacuum Catastrophe'' reflects the fundamental fact that the
quark-gluon condensate (responsible for the color confinement in
QCD) is off by over forty orders of magnitude and has a wrong sign
with the observable $\Lambda$-term value, whereas the classical
weakly-coupled Higgs condensate (responsible for the electroweak
symmetry breaking in the Standard Model) is off by over fifty orders
of magnitude while having the correct sign. In one way or another,
the ``exact compensation hypothesis'' should be realized in Nature
in order to explain the strong suppression of the observable Dark
Energy density compared to that of the individual condensates
contributions to the ground state energy of the Universe. The exact
compensation of weakly-coupled perturbative vacua (e.g. the Higgs
condensate) components at short distances (essentially by virtual
particles in perturbative loops) is typically assumed to take place
in a high-scale grand-unified and/or superstring-inspired ``Theory
of Everything'' and is not a subject of our present work. In this
work we discuss a possible compensation of the largest
strongly-coupled non-perturbative component of the ground state
energy, the quark-gluon condensate, which is the lowest-energy and
the most troublesome one in Particle Physics.

Specifically, the physical and mathematical premises for the first
condition above are contained in physically reasonable estimates for
the energy density possibly coming from the finite-time instanton
solutions of the cosmological Yang-Mills fields as discussed in
Ref.~\cite{Pasechnik} and/or from vacuum hadronic wave modes given
by Eq.~(\ref{h}). Both of these estimates show that the respective
contributions compensating a large negative instanton energy
(\ref{top}), may have a similar nature being components of the same
{\it non-perturbative QCD vacuum}. An interesting analogy of the
co-existence of vacuum contributions of essentially different types,
but of the same origin, takes place e.g. in solid state physics: in
a crystallization process there are both negative (binding energy of
atoms in the crystal lattice) and positive (zero-point collective
fluctuations of the lattice itself) contributions to the ``vacuum''
energy of the medium. Of course, such an analogy is not exact and
there is no exact compensation of these contributions in this case.

The problem of exact compensation of vacuum contributions at the QCD
energy scale seemingly has a dynamical nature. No doubts, the QCD
vacuum state in the modern Universe has been created during its
cosmological evolution: the quantum-topological vacuum structures
(e.g. instantons) have been created in real time around the
quark-hadron phase transition in the Universe evolution. The quantum
state of vacuum collective (wave) fluctuations of such structures
should, therefore, be matched to the quantum state of these
structures themselves. At the same time, the unprecedented accuracy
in cancellation between different vacuum contributions to many
relevant digits (or their major fine tuning), if it takes place in
Nature, forces us to consider such a matching of the vacuum
topological structures and their collective fluctuations as a {\it
new physical phenomenon}. The latter phenomenon must be generic for
all existing vacua and, supposedly, can provide a key to deeper
understanding of the color confinement problem. Most probably, an
adequate description of this phenomenon is the subject of a new
dynamical theory of the non-perturbative QCD vacuum which has not
been yet created. However, we have shown that the compensation of
the quantum-topological and quantum-wave contributions at
macroscopic length scales is phenomenologically motivated. In a
sense, such an compensation is a manifestation of the QCD
confinement since there are practically no non-zeroth gluon fields
propagating at the length scales larger than the typical hadron
scale $\sim1$ fm or so, and certainly disappear at macroscopically
large cosmological scales typical for modern Universe.

The second condition above is tightly connected to the first one.
Indeed, as is seen from Eqs.~(\ref{xx'}), (\ref{top1}) and
(\ref{d}), the representation of the two-point function $D(x-x')$ as
a difference between topological and hadron correlation functions
appears as a consequence of our assumption about the exact
compensation of the topological and hadron wave contributions to the
vacuum energy density calculated without the graviton-mediated
corrections. As we will see below, the smallness of the QCD
parameter $\Delta \sim 3\cdot 10^{-6}$ introduced in Eq.~(\ref{Del})
is required by phenomenological arguments -- both by data on the
$\Lambda$-term density itself and by QCD phenomenology. On the other
hand, the parameter $\Delta$ is a functional of the difference
between the correlation functions defined at the same space-time
scales (\ref{sc}). Thereby, there are physical and mathematical
premises for the mutual cancellation of topological and wave
contributions also in the effect graviton-mediated interactions, but
such a cancellation is not exact. Thus, without a self-consistent
theory of the non-perturbative QCD vacuum the actual value of the
residual effect and, hence, the genuine value of $\Lambda$-term
density can only be estimated by means of available experimental
data and phenomenological arguments only. Based on purely
phenomenological QCD arguments, let us try to move on towards a
simple numerical estimation of the $\Lambda$-term density starting
from Eq.~(\ref{fin}).

Indeed, the small parameter $\Delta \sim 3\cdot 10^{-6}$ can, in
principle, be expressed through the well-known QCD parameters.
Substituting only topological $D_{top}(x-x')$ or hadronic
$D_{h}(x-x')$ part of the complete correlation function $D(x-x')$
into Eq.~(\ref{Del}) one immediately obtains $\Delta \sim 1$.
Meanwhile, it is known that naturally small QCD parameters are the
light quark $u,d,s$ masses, so let us assume that the value of
(\ref{Del}) with the complete QCD correlation function $D(x-x')$ is
different from zero due to the {\it chiral symmetry breaking}
effects. A small shift between the characteristic scales of
topological and hadronic fluctuations induced by the chiral symmetry
breaking can by given in terms of small current quark
$m_u,\,m_d,\,m_s$ masses
\[
\begin{array}{c}
\displaystyle 1/L_{top}\sim 1/L_h\sim 1/L_g\,,
\\[3mm]
\displaystyle |1/L_{top}-1/L_h| \sim m_u+m_d+m_s\,,
\end{array}
\]
leading to a good estimate
\begin{equation}
\displaystyle \Delta = k\cdot
\frac{(m_u+m_d+m_s)^2L_g^2}{(2\pi)^4}>0\,, \label{Del1}
\end{equation}
where $k\sim 1$ is the dimensionless factor, and the factor
$1/(2\pi)^4$ appears in the Fourier transform of the corresponding
Green function. Then, the experimentally observable value of the
$\Lambda$-term density (\ref{Lambda-mpi}) can be naturally obtained
for $k=1$ and the light quark masses satisfying the approximate sum
rule $m_u+m_d+m_s\simeq 100\;\text{MeV}$, which is in agreement with
experimentally known values
$m_u=1.5\;-\;5\;\text{MeV},\;m_d=3\;-\;9\;\text{MeV},\;m_s=60\;-\;170\;\text{MeV}$.
In this case, the phenomenological formula for the $\Lambda$-term
density, obtained based upon Eqs.~(\ref{fin}) and (\ref{Del1}),
contains all the basic QCD and chiral symmetry breaking parameters,
i.e.
\begin{equation}
\begin{array}{c}
\displaystyle \varepsilon_{\Lambda}=\pi G\langle
0|:\frac{\alpha_s}{\pi}F_{ik}^aF_a^{ik}:|0\rangle^2\times\left(\frac{b_{\rm
eff}}{8}\right)^2 \frac{(m_u+m_d+m_s)^2L_g^4}{(2\pi)^4}
\ln\frac{L_{g}^{-1}}{e\Lambda_{\rm
QCD}}\ln\frac{L_{g}^{-1}}{\Lambda_{\rm QCD}}\,.
\end{array}
\label{fin1}
\end{equation}
In fact, Eq.~(\ref{fin1}) provides a naive estimate for the
dynamical QCD-induced {\it positive} contribution to the
$\Lambda$-term energy density $\varepsilon_{\Lambda}>0$ which is
close to the observed value within a factor of few. Quantitatively,
the size of this dynamically induced $\Lambda$-term can be sensitive
to details of yet unknown non-perturbative QCD dynamics. For a
better estimate, one would certainly like to evaluate the
$\Delta$-factor in Eq.~(\ref{fin}) to a higher precision
incorporating the chiral symmetry breaking effects e.g. in the
lattice QCD framework or elsewhere.

The applicability of both conditions indicated above and hence the
quantitative conclusions of the paper actually extend only up to the
QCD phase transition epoch which has likely happened at temperatures
of the Universe of the order of $\sim100$ MeV. Above this
temperature quarks and gluons in the cosmological plasma do not form
bound states, hadrons, and the medium is in the phase of the
quark-gluon matter. Also, at higher temperatures, there is no quark
condensate, i.e., the phase of unbroken chiral symmetry is realized.
This basically means that at $T>100$ MeV there was no any
compensation mechanism of the non-perturbative quantum-topological
(instanton) quark-gluon contribution like the one discussed in the
Section II of this paper since the quantum-wave (hadronic) modes
were completely ``melted up'' at these temperatures in the
deconfined phase of the cosmological plasma; those appear only when
the Universe gets cooler than the QCD phase transition temperature.
So, at $T>100$ MeV both conditions are not valid any longer (the
basic analytic result of the paper given by Eq.~(\ref{laint1})
remains always valid, of course, irrespectively of any of the above
physical conditions).

Therefore, the energy density of the Universe during earlier pre-Big
Bang epochs (e.g. inflation) was utterly dominated by the
uncompensated quantum-topological QCD contribution given by
Eq.~(\ref{top}) and possibly by other perturbative vacua at $T>100$
GeV, which are larger compared to the gravity-induced correction
(\ref{laint1}) (with only quantum-topological correlation function
$D_{top}$ incorporated) at least by 38 orders of magnitude in
absolute value or more. In this sense, the gravity-induced
correction given by Eq.~(\ref{laint1}) cannot be relevant for the
inflationary stage in the Universe evolution -- it becomes tiny
there compared to other vacua contributions. The same conclusion
about the strong dominance of the initially very large
(uncompensated) vacuum density over the graviton-induced corrections
to it applies to any inflationary scenario since below the Planck
scale the graviton-induced corrections are generally very small and
strongly suppressed compared to other vacua. These corrections
become important only after the QCD phase transition as that is the
only part of the vacuum energy density which remains uncompensated
at macroscopically large length scales. A more detailed analysis of
the early pre-Big Bang epochs in relation to these aspects including
particle production mechanisms after inflation is required and would
certainly be a plausible direction for further studies.

To conclude, we notice that under the above two conditions the
$\Lambda$-term generation at the QCD phase transition scale appears
to be a natural phenomenon for supersymmetric (or superstring)
scenarios of high-energy physics. Indeed, within these scenarios the
vacuum subsystem of a quantum-topological nature exists only in the
QCD sector which is the lowest one among the energy scales in
Nature. All other vacuum subsystems consist mainly of perturbative
(weakly-coupled) vacuum fluctuations of fundamental fields and one
may expect that e.g. a proper supersymmetry theory is capable of
explaining the exact compensation of their contributions into the
$\Lambda$-term density if $\sum_{\rm bos} M_{\rm bos}^6\equiv
\sum_{\rm ferm} M_{\rm ferm}^6$ is satisfied. This can be seen e.g.
from Eq.~(\ref{Lambda-mpi}) where the respective energy scale
(virtual particle mass of bosons $M_{\rm bos}$ and fermions $M_{\rm
ferm}$) appears to the sixth power while bosons and fermions
contribute to the vacuum energy with opposite signs. A detailed
analysis of cancellations of this type and their cosmological
consequences is planned for a forthcoming study.

{\bf Acknowledgments}

Stimulating discussions and correspondence with Sabir Ramazanov,
Johan Rathsman and Torbj\"orn Sj\"ostrand are gratefully
acknowledged. This work is supported in part by the Crafoord
Foundation (Grant No. 20120520).

\end{document}